\documentclass[12pt]{iopart}
\usepackage{setspace}
\usepackage{color}
\usepackage{graphicx}
\renewcommand{\deg}{$^\circ$}
\newcommand{\muH}{$\textrm{$\mu$}_0 H$}
\newcommand{\Hpc}{$H\|$c~}
\newcommand{\Hpab}{$H\|$a,b~}
\newcommand{\HpcTemp}{$H\|$c}
\newcommand{\HpabTemp}{$H\|$a,b}
\begin{document}

\title[Stress dependence of $I_\mathrm{c}$ in neutron irradiated coated conductors]{Stress dependence of the critical currents in neutron irradiated (RE)BCO coated conductors}
\date{1$^{st}$ of June 2012}
\author{J Emhofer, M Eisterer, and H W Weber}

\address{Atominstitut, Vienna University of Technology, Stadionallee 2, 1020 Vienna, Austria}
\ead{eisterer@ati.ac.at}

\begin{abstract}
The application of HTS coated conductors in future fusion or accelerator magnets is currently of increasing interest. High Lorentz forces and therefore high hoop stresses act on the conductors in large coils.
The conductor is furthermore exposed to neutron radiation in fusion or accelerator magnets. The expected neutron fluence over the desired lifetime of such magnets can be simulated by irradiation experiments in a fission reactor. The coated conductors were characterized in the pristine state and after irradiation to a fast neutron fluence of $1\,\times$\,$10^{22}$\,m$^{-2}$\,(ITER design fluence). The sensitivity of the critical currents to applied stress was measured in liquid nitrogen. The cold part of the set-up was positioned between a rotatable split coil electro-magnet for assessing the $I_\mathrm{c}$-anisotropy up to 1.4\,T under maximum Lorentz force configuration.

The $I_\mathrm{c}$-sensitivity to applied stress changed significantly in the GdBCO/IBAD-conductors after irradiation, whereas nearly no change was observed in the YBCO/RABiTS-conductor. Furthermore, $I_\mathrm{c}$ and $T_\mathrm{c}$ were strongly reduced in the GdBCO/IBAD-sample after irradiation. The angular dependence of $I_\mathrm{c}$ changed for both samples in different ways after the irradiation, but no change in the angular dependence was observed upon applying stress.

The high neutron capture cross-section of Gd and the resulting strong reduction of $T_\mathrm{c}$ seem to be responsible for the different stress dependence of $I_\mathrm{c}$ in irradiated Gd-123 coated conductors.
\end{abstract}

\maketitle

\section{Introduction}
Impressive progress in coated conductor development has been made in recent years\,\cite{HolesingerAM20}-\nocite{RupichSUST23} \cite{SelvamanickamIEEE19}. The possibility of using liquid nitrogen as the coolant, the high mechanical stability\,\cite{VanDerLaanIEEE22}\nocite{VanDerLaanSUST23}\nocite{CheggourIEEE15}\nocite{UgliettiSUST19}\nocite{HigginsIEEE21}-\nocite{OsamuraSUST22}\cite{SuganoSUST21} and the robustness of YBCO coated conductors against neutron irradiation\,\cite{EistererSUST23}\cite{FugerIEEE19} render YBCO coated conductors promising candidates for applications in future fusion or accelerator magnets. Tensile stress or strain affect the performance of the conductors reversibly up to the irreversible stress/strain limit\,($\sigma_\mathrm{irr}$ / $\varepsilon_\mathrm{irr}$). In the case of Bi-2223 conductors the stress-dependence of $I_\mathrm{c}$ was entirely explained by the pressure dependence of $T_\mathrm{c}$\,\cite{VanDerLaanSUST24-3}. However, since the pressure dependence of $T_\mathrm{c}$ is highly anisotropic in YBCO\,\cite{WelpPRL69}, the manufacturing method and in particular the alignment of the unit-cells strongly affect the stress dependence of $I_\mathrm{c}$ in YBCO. Recent studies \cite{VanDerLaanSUST24-3}\cite{VanDerLaanIEEE22} demonstrated that the reversible stress/strain-effects in YBCO/MOCVD and DyBCO/ISD conductors are also determined by the uni-axial pressure dependence of $T_\mathrm{c}$.
Furthermore, a strong field dependence of the $I_\mathrm{c}$ sensitivity on the applied tensile and/or compressive strain was found for YBCO/IBAD/MOCVD\,\cite{VanDerLaanIEEE22}\cite{VanDerLaanSUST23}\cite{HigginsIEEE21}, YBCO/RABiTS/MOD\,\cite{CheggourIEEE15}\cite{UgliettiSUST19} and YBCO/MOCVD/PLD/IBAD\,\cite{SuganoSUST21} samples. 

The change in $I_\mathrm{c}$ caused by neutron radiation depends on the superconducting material, the neutron energy, magnetic field and temperature. Fast neutrons (E\,$>$\,0.1\,MeV) generate collision cascades in cuprates by transferring sufficient energy to a primary knock-on atom, which initiates further collisions \cite{FrischherzPhysicaC232}\cite{AleksaPhysicaC297}. These collisions lead to local melting of the lattice and the formation of spherical defects with a diameter of $\approx$ 6\,nm \cite{FrischherzPhysicaC232}. The size of these normal conducting impurities matches the size of a flux line core (2$\xi_{a,b} \approx 6\,$nm at 77\,K) and is, therefore, optimal for flux pinning. Furthermore, these defects are uncorrelated and randomly distributed. Fast-neutron-induced cascades have been shown to be responsible for flux pinning and intra-grain critical current enhancement in (RE)BCO \cite{SickafusPRB46}\cite{FrischherzPhysicaC232}. 

Thermal and epi-thermal neutrons do not lead to extended defects in YBCO. Epi-thermal neutrons\,(in the keV range) may create point defects or clusters, if the recoil energy of the primary knock-on atom is high enough to displace at least one atom. These point defects disturb the regularity of the CuO$_2$ planes,\cite{ChudySUST25}, which reduces the critical temperature. The reduction in $T_\mathrm{c}$ is rather weak in YBCO \,(about -2\,K for a fast neutron fluence $\Phi \cdot t = 10^{22}\,$m$^{-2}$ \cite{EistererSUST23}, where $\Phi$ and $t$ denote the fast neutron flux density and the irradiation time, respectively). Thermal neutrons\,($E<0.5$\,eV) do not affect the material by direct collisions, as the energy transferred by elastic collisions is below the displacement energy of a single atom. The neutron capture cross is negligible in YBCO. 

Nuclei with high neutron capture cross-sections\,$\sigma(n,\gamma)$ should be avoided when the materials have to be exposed to neutron irradiation. Unfortunately, state-of-the-art (RE)BCO coated conductors contain gadolinium\,(GdBCO) or samarium\,(SmBCO) having high neutron capture cross-sections. The thermal neutron capture cross-sections of the two stable isotopes $^{155}$Gd and $^{157}$Gd with a natural abundance of 14.8\% and 15.7\%, respectively, are as high as $6.074 \times 10^4$\,b and $2.537 \times 10^5$\,b, respectively \cite{nndc}. (One barn, b, corresponds to $10^{-28}$\,$m^{2}$.) After capturing a neutron, the excited nucleus emits $\gamma$-rays and the corresponding recoil energy is transferred to the nucleus. This recoil is sufficiently high to displace the excited $^{156}$Gd\, and $^{158}$Gd nuclei thus creating a point defect. Neglecting the kinetic energy of the neutron, the recoil energy was calculated to be 29\,eV and 34\,eV for the excited $^{156}$Gd and $^{158}$Gd nuclei\,\cite{SickafusPRB46}. Hence, in Gd or Sm-compounds the thermal and epi-thermal neutrons of the TRIGA reactor spectrum play a crucial role and should be removed for the simulation of a fusion spectrum by a fission reactor, since the expected fluence of low energy neutrons is very small at the magnet location of a fusion device. 

The results reported in this work refer to GdBCO/IBAD and YBCO/RABiTS conductors irradiated with the entire neutron energy spectrum of the TRIGA reactor\,\cite{WeberJNM137}.

\section{Samples}
Two different (RE)BCO coated conductors were used for this study. The SuperPower SCS4050 tape, which consists of a GdBa$_2$Cu$_3$O$_{7-\delta}$ (GdBCO) superconducting layer on an IBAD\,(Ion Beam Assisted Deposition) template and the AMSC 344C\,(Amperium) tape consisting of an YBa$_2$Cu$_3$O$_{7-\delta}$ (YBCO) superconductor on a RABiTS\,(Rolling Assisted Bi-axially Textured Substrate) template.

The GdBCO layer in the GdBCO/IBAD sample was grown by MOCVD\,(Metal Organic Chemical Vapour Deposition) on a Hastelloy substrate with a thickness of 50\,$\mu$m. On top of the GdBCO-layer a 2\,$\mu$m silver layer was deposited and the entire sample was surrounded by 20\,$\mu$m copper. The individual samples were 50\,mm long, 4.04\,mm wide and 0.1\,mm thick.

The \emph{RE}BCO layer in the YBCO/RABiTS contains yttrium and dysprosium with a nominal composition of Y:Dy:Ba:Cu of 1:0.5:2:3. The tape was grown by MOD\,(Metal Organic Deposition) onto a sputtered buffer stack. The cold worked and annealed (RABiTS) substrate of this tape was a 75\,$\mu$m thick Ni-5 at\% W alloy. A $\sim$2\,$\mu$m silver layer was deposited on top of the YBCO layer and the entire tape laminated with a copper foil for better handling and stabilization. The samples were 50\,mm long, 4.4\,mm wide and 0.21\,mm thick.

Note that different samples were used for the characterization in the pristine and the irradiated state, marked by \#1 and \#2 in the figures, respectively. Both samples were cut from the same spool and the self-field $I_\mathrm{c}$ at 77\,K as well as the homogeneity of the \#2-samples were checked before the irradiation by four point measurements and magnetoscans, respectively. As the YBCO/RABiTS\#2 was irreversibly damaged before reaching the irreversibility limits, a third YBCO/RABiTS sample denoted by \#3 was used exclusively for a self-field $I_\mathrm{c}$ measurement in the unirradiated state. Furthermore, shorter reference samples of 26\,mm length from the same spools were irradiated in the same capsule for measurements of the irreversibility line. 

\section{Experimental Details}
\subsection{Tensile stress set-up}
\begin{figure}[htpb]
	\centering
	\begin{center}
		\includegraphics[width=4cm]{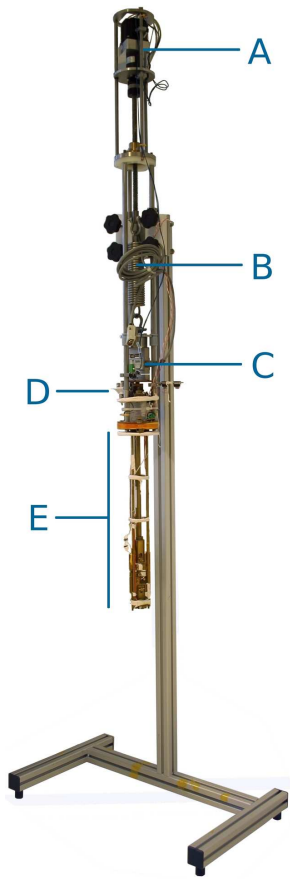}
		\caption{\label{machine}Photograph of the tensile stress insert. (A)\,Stepper motor, (B)\,removable spring, (C)\,load cell, (D)\,connections, heating resistor, thermal switch, safety spring, etc., (E)\,cold part.}
	\end{center}
\end{figure}

The cold part of our tensile stress insert\,(E in figure \ref{machine}) was positioned between a rotatable split coil electro-magnet. Horizontal fields of up to 1.4\,T can be applied at different orientations\,($\Theta$). The angular range assessed in our experiments was about 220\deg , including both a,b-peaks, in angular steps of 1\deg ~around $H\|$c and $H\|$a,b and 3\deg ~elsewhere.
The cold part was immersed in liquid nitrogen during the measurements. The temperature variation of liquid nitrogen due to changes of ambient pressure or oxygen pick-up induces some scatter in the data. To reduce this error to an acceptable level, all data were discarded whenever the temperature at the sample position was not within $77.4\pm$0.3\,K during the measurement.

\begin{figure}[htpb]
	\centering
	\begin{center}
		\includegraphics[width=7cm]{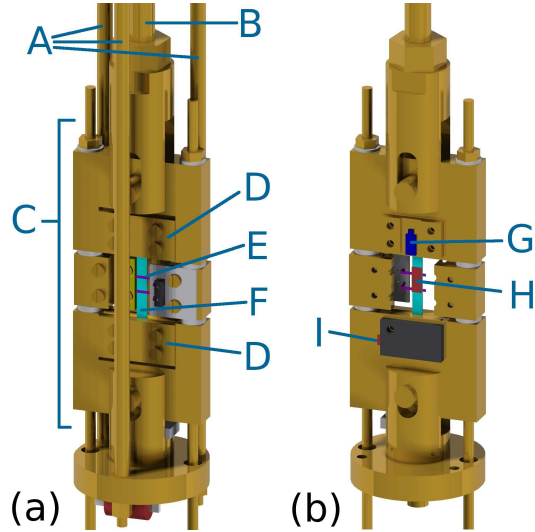}
		\caption{\label{SetUpSolidEdge}CAD-drawings of the lower cold part of the set up. The front side is illustrated in (a) whereas the back side, without the push rods and with an artificial cut-out in the centre, is illustrated in (b). Wires and cables are not depicted. (A)\,Push rods, (B)\,pull rod, (C)\,removable carriage, (D)\,current clamps, (E)\,voltage contacts, (F)\,coated conductor, (G)\,Hall probe, (H)\,strain gauge, (I)\,reference strain gauge on a stress-free reference coated conductor.}
	\end{center}
\end{figure}

Tensile forces were set with a stepper motor\,(A in figure \ref{machine}). A trapezoidal thread transformed the rotation to a translation along the pull rod axes. The force was transduced to the pull rod\,(B in figure \ref{SetUpSolidEdge}) by a 1000\,kN spring\,(B in figure \ref{machine}) and a ball bearing prevents the lower parts from rotating. A 2000\,kN load cell\,(C in figure \ref{machine}) located between the ball bearing and the pull rod, measured the applied force.

Figure \ref{SetUpSolidEdge} shows the front and back side of the cold part. The coated conductor\,(F in figure \ref{SetUpSolidEdge}) is clamped between the upper and the lower part of the removable carriage\,(C in figure \ref{SetUpSolidEdge}). The pull and the push rods also act as current bars and therefore brass was used for the cold part. The upper and lower parts of the removable carriage are insulated from each other by Teflon inserts. A Pt100-Class 1/5 sensor was used to monitor the temperature close to the sample. 

The voltage contacts\,(E in figure \ref{SetUpSolidEdge}) were glued with silver paste, allowing small position rearrangements due to elongation. The average distance between the contacts was 5\,mm and $I_\mathrm{c}$ was evaluated using a 1\,$\mu$V/cm criterion by fitting the data by a power law\,($E=E_c\,(I/I_c)^n$, least squares method). 

A strain gauge was glued directly onto the coated conductor with epoxy resin\,(H in fig. \ref{SetUpSolidEdge}), measuring the strain of the sample. A reference strain gauge was mounted close to the sample\,(I in figure \ref{SetUpSolidEdge}) on a small stress free coated conductor of the same type as the measured sample in order to subtract the strain caused by cooling the sample to liquid nitrogen temperatures.  Both strain gauges were connected to a tunable Wheatstone bridge.

A Hall probe was mounted parallel to the tape surface in order to measure the orientation of the applied magnetic field, which was set by rotating the split coil with a stepper motor.

The full $I_\mathrm{c}$-anisotropy characterization at a fixed applied force was done during one measurement cycle.
Such a cycle started by increasing the force stepwise from the relaxed state to the desired force. At each step, the self-field $I_\mathrm{c}$ was measured. Between each step, the stress was released and $I_\mathrm{c}$ was re-measured. The dependence of the self-field $I_\mathrm{c}$ on the stress/strain was obtained from these measurements. The anisotropy measurements started after reaching the desired set-force. Angular resolved $I_\mathrm{c}$-measurements were performed at 0.1\,T, 0.2\,T, 0.4\,T, 0.6\,T, 1\,T and 1.4\,T. After the last measurement, the computer program started the warm-up process. 
Typically, anisotropy measurements were performed at nine different forces, leading to a net measurement time of typically around 400\,hours for one sample including periods of heating to room temperature. Note, that in the ``relaxed'' state a force of 10\,N\,(11\,MPa in YBCO/RABiTS and 25\,MPa in GdBCO/IBAD) was applied to the samples.

\subsection{Neutron irradiation}
The samples were irradiated in the central irradiation facility of the TRIGA Mark II research reactor in Vienna to a fast neutron fluence of $\Phi \cdot t$ = $1\,\times$\,$10^{22}$\,m$^{-2}$\,($E >0.1$\,MeV). The samples were sealed in a quartz tube and the temperature did not exceed 70\deg\,C during the irradiation. The introduced defect structure in single-crystalline YBCO was already studied in detail \cite{SauerzopfPhysRevB57} and further irradiation studies of coated conductors can be found in \cite{FugerIEEE19}. 

\subsection{Magnetoscan}
The homogeneity of the superconducting layer was checked by magnetoscan imaging before and after applying the maximum tensile stress. A detailed description of the set-up can be found in \cite{EistererSUST16}. The local magnetic field detected slightly above the sample but below a permanent magnet is closely related to the local current density and, therefore, provides evidence for the local quality of the sample \cite{ZehetmayerAPL90}.

\section{Results and Discussion}
\subsection{Stress - strain dependence}
\begin{figure}[htpb]
	\centering
	\begin{center}
		\includegraphics[width=7cm]{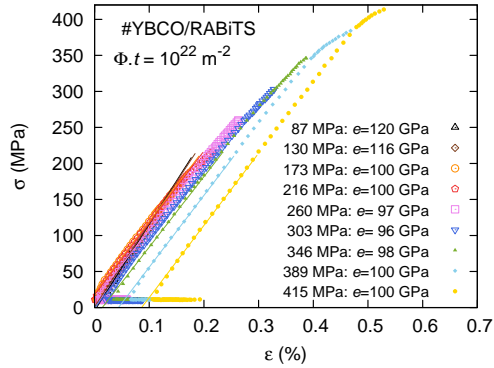}
		\caption{\label{StressStrainIrradiatedYBCO}Stress-strain dependence of the irradiated YBCO/RABiTS sample. Different symbols denote different maximum applied stresses.}
	\end{center}
\end{figure}

The irradiated YBCO/RABiTS sample (figure \ref{StressStrainIrradiatedYBCO}) was stressed up to the irreversible limits $\sigma_{irr}=$410\,MPa and $\varepsilon_{irr}=$0.52\%. 
The yield strength of the substrate was reported to be 260\,MPa\,\cite{Ekin} at 76\,K which corresponds to a stress of 84\,MPa in the YBCO/RABiTS tape. The applied stress led to a distinct irreversible deformation at the strain gauge position after applying 346\,MPa (cf. figure~\ref{StressStrainIrradiatedYBCO}). A reduction of the irreversible deformations was always observed after heating the sample to room temperature.

The Young's modulus\,($e$) was evaluated by fitting the data between 20\,N\,(22\,MPa in YBCO/RABiTS and 50\,MPa in GdBCO/IBAD) and the maximum applied stress of the previous measurement loop (e.g. 173\,MPa for the 216\,MPa measurement). The values found for the irradiated YBCO/RABiTS sample are quoted in figure \ref{StressStrainIrradiatedYBCO}.

No change of Young's modulus was observed in the YBCO/RABiTS tape after irradiation, but a slight decrease by 24\,GPa was observed in the GdBCO/IBAD conductor. Since these results represent the first measurements ever on irradiated GdBCO/IBAD coated conductors and in view of the lack of sufficient statistics, further experiments will be needed for confirmation.
The presently available average values of the Young's modulus are listed in table \ref{StressStrainValues}.

\subsection{Self-field $I_\mathrm{c}$}
Self-field $I_\mathrm{c}$ measurements were performed together with the stress-strain measurements (except for the YBCO/RABiTS\#3 sample). Figures \ref{SelfFieldIcYBCO} and \ref{SelfFieldIcGdBCO} show the data up to the highest applied stress, where an irreversible reduction of $I_\mathrm{c}$ was observed for the first time. The critical currents are normalized by the mean value (before the samples were irreversibly damaged) of the critical currents in the relaxed state and at zero field. (A small force of 10\,N was applied also in the relaxed state in order to straighten the tape. Because of the different geometric cross sections of the tapes, this corresponds to a different pre-stress of 11\,MPa in the YBCO/MOD tape and of 25\,MPa in the GdBCO/IBAD tape.) The irreversible stress/strain limit was defined as the last value of applied stress/strain where $I_\mathrm{c}$ in the (following) relaxed state was above 95\,\% of the initial value.

\subsubsection{YBCO/RABiTS}

\begin{figure}[htpb]
	\centering
	\begin{center}
		\includegraphics[width=7cm]{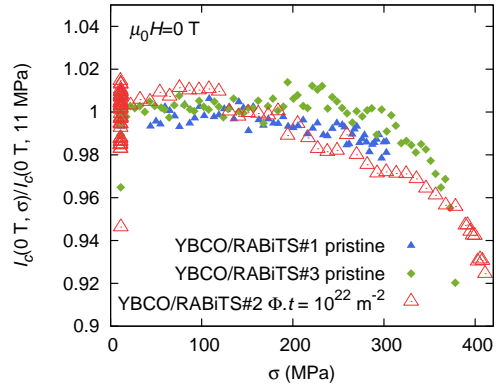}
		\caption{\label{SelfFieldIcYBCO}Self-field $I_\mathrm{c}$ dependence on stress\,(YBCO/RABiTS).}
	\end{center}
\end{figure}

The stress dependence of $I_\mathrm{c}$ in the YBCO/RABiTS conductor did not change significantly after irradiation (figure \ref{SelfFieldIcYBCO}). 
However, it seems that a small maximum appears at around 100\,MPa after irradiation. The pristine YBCO/RABiTS\#1 sample was irreversibly damaged during the warm-up process after applying 303\,MPa, therefore no data exist for higher stress values. Below 303\,MPa\,($\varepsilon \approx $0.33\%), $I_\mathrm{c}$ was never reduced to below 98\% of the original $I_{c}$. In the YBCO/RABiTS\#3 sample, the irreversible stress limit was reached at 373\,MPa\,($\varepsilon \approx$ 0.46\%). $I_\mathrm{c}$ of the irradiated sample was reduced to 92\,\% just below the irreversible stress limit of 410\,MPa\,($\varepsilon \approx$ 0.53\%).

The relaxed $I_{c}$ decreases after irradiation to 76\% of $I_\mathrm{c}$ in the pristine sample. A $T_\mathrm{c}$ reduction of 1.8\,K was measured in the reference sample. Furthermore, $B_{irr}(77\,K)$ was only slightly enhanced at this fluence. 

\subsubsection{GdBCO/IBAD}
\begin{figure}[htpb]
	\centering
	\begin{center}
		\includegraphics[width=7cm]{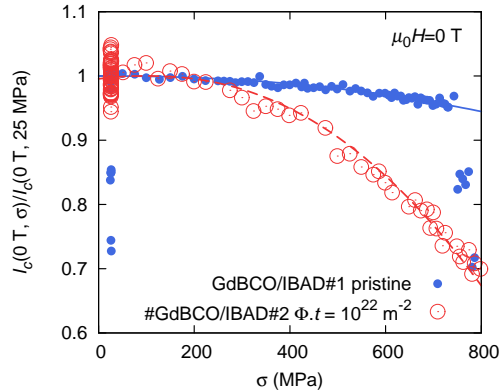}
		\caption{\label{SelfFieldIcGdBCO}Self-field $I_\mathrm{c}$ dependence on stress\,(GdBCO/IBAD).}
	\end{center}
\end{figure}

Figure \ref{SelfFieldIcGdBCO} shows the $I_\mathrm{c}$-stress dependence of the GdBCO/IBAD sample.

After applying 740\,MPa\,($\varepsilon \approx $0.55\%), the critical current was permanently reduced in the pristine sample. Two major reductions occurred between 740\,MPa and 800\,MPa. After the first reduction, $I_{c}$ in the relaxed state was reduced to $\approx$83\% of its initial value. After applying 790\,MPa, $I_{c}$ in the relaxed state was reduced to $\approx$70\% of the initial value. Just below the irreversible stress limit of 740 MPa, $I_\mathrm{c}$ was reduced only by 5\%. Note that measurements of other samples from the same spool indicated irreversible stress limits of up to 800 MPa.

For the irradiated GdBCO/IBAD sample, the irreversible stress/strain limit was reached at 800\,MPa/$\varepsilon$=0.65\,\%. $I_\mathrm{c}$ of the irradiated sample is more sensitive to applied stress than in the pristine sample. Close to the irreversible stress limit, $I_\mathrm{c}$ was reduced to 67\% of the relaxed value.

The relaxed self-field $I_{c}$ was reduced by one order of magnitude after irradiation. $T_\mathrm{c}$ in the reference sample was reduced by 6.2\,K and also the irreversibility field significantly decreased at this temperature.

The lower $T_\mathrm{c}$ and the resulting higher $T/T_c$ ratio of the irradiated tape at 77\,K seem to be responsible for the change of the $I_\mathrm{c}$-sensitivity to applied stress. A higher $I_\mathrm{c}$-sensitivity at temperatures closer to $T_\mathrm{c}$ was found for YBCO tapes in Ref. \cite{SuganoSUST21} and is expected from the equation for a single grain \cite{VanDerLaanSUST24N11}:
\begin{equation}
 I_{c,sg}(T,\sigma,b)=I_c(T=0,\sigma=0)\left(1-\frac{T}{b\,\sigma+T_{c}(0)}\right)^{1.5}
\label{EquIc}
\end{equation}
where $T_c(0)$ is the critical temperature at zero applied strain and $b$ denotes the change of $T_\mathrm{c}$ with applied pressure. The anisotropic pressure dependence of $b$ was found to be $b_a$=-2.0$\pm$0.2\,K/GPa and $b_b$=1.9$\pm$0.2\,K/GPa with pressure along the a-axis and the b-axis, respectively \cite{WelpPRL69}. For small $\sigma$ equation~\ref{EquIc} leads to first order to 
\begin{equation}
I_\mathrm{c}(\sigma)/I_\mathrm{c}(0)=1-1.5\frac{bT}{T_\mathrm{c}(T_\mathrm{c}-T)}\sigma + o(\sigma^2).
\end{equation}
The strain sensitivity thus increases with decreasing transition temperature, in particular, if $T_\mathrm{c}$ approaches $T$, as in the irradiated GdBCO/IBAD sample. Note that a pronounced curvature in $I_\mathrm{c}(\sigma)$ is observed experimentally instead of the predicted linear behavior. This is a consequence of the dependence of the critical currents on the applied electric field.  
A model with two grains oriented with their a-axis parallel and perpendicular to the tape direction was presented in \cite{VanDerLaanSUST24N11}, which describes the stress dependence of $I_\mathrm{c}$ in YBCO coated conductors qualitatively. However, since a real coated conductor is a percolative network of differently aligned grains, a quantitative model needs to be developed for a comparison with experimental data.

\begin{table}
\caption{\label{StressStrainValues}Summary of the parameters obtained from the self-field measurements. ($T_c$ from reference samples)}
\begin{indented}
\item[]\begin{tabular}{@{}lllllll}
\br
Sample & $\sigma_\mathrm{irr}$ & $\epsilon_{irr}$ &$T_c$& $I_{c}(0$\,T,relaxed)&n-value&$\overline{e}$ \\
\mr
pristine YBCO\#1& $>$303\,MPa & $>$0.33\,\% &89.8\,K &88.4\,A&30.0&106\,GPa\\ 
pristine YBCO\#3& 373\,MPa & $>$0.46\,\% &89.8\,K &93.6\,A&30.7&115\,GPa\\ 
irradiated YBCO\#2& 410\,MPa & 0.53\,\% &88.0\,K &67.1\,A&25.8&103\,GPa\\ 
pristine GdBCO\#1& 740\,MPa & 0.55\,\% & 92.9\,K&142.7\,A&33.2&163\,GPa\\ 
irradiated GdBCO\#2 & 800\,MPa & 0.65\,\%& 86.7\,K& 16.52\,A&18.2&139\,GPa\\ 
\br
\end{tabular}
\end{indented}
\end{table}

\subsection{Homogeneity}
\begin{figure}[htpb]
	\centering
	\begin{center}
		\includegraphics[width=8cm]{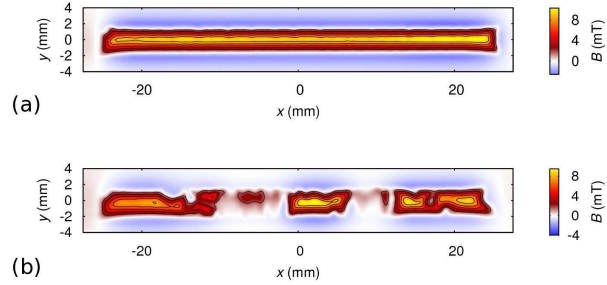}
		\caption{\label{Homogeneity}Magnetoscan images of the pristine GdBCO\#1 sample before\,(a) and after\,(b) applying a tensile load of 320\,MPa.}
	\end{center}
\end{figure}

Figure \ref{Homogeneity} shows magnetoscan images of the pristine GdBCO sample before and after the tensile stress measurements.
The virgin conductor\,(figure \ref{Homogeneity}a) is very homogeneous without significant defects. The conductor was irreversibly damaged after applying the maximum load of 740\,MPa. After removing the sample from the set-up, no destruction was found by optical inspection. The magnetoscan image\,(figure \ref{Homogeneity}b) reveals the degree of damage. The superconducting layer was damaged over a long length and not only along single distinctive macroscopic cracks. The two main defects around $x=-5\,$mm and $x=10\,$mm may correlate with the two significant reductions of $I_\mathrm{c}$ at $\approx$740\,MPa and $\approx$800\,MPa presented in figure \ref{SelfFieldIcGdBCO}. 

Note that the left and right ends, where the current clamps held the sample, are less damaged. Hence, the pressure exerted by the current terminals does not destroy the sample and a good current feed is ensured.

\subsection{In-field $I_\mathrm{c}$ anisotropy}

The angular dependence of $I_\mathrm{c}$ at liquid nitrogen temperature changes completely after irradiation in both samples.

In the YBCO/RABiTS conductor the strong additional pinning centres introduced by fast neutron irradiation lead to an enhancement of $I_\mathrm{c}$ between the two a,b peaks. 

\begin{figure}[htpb]
	\centering
	\begin{center}
		\includegraphics[width=7cm]{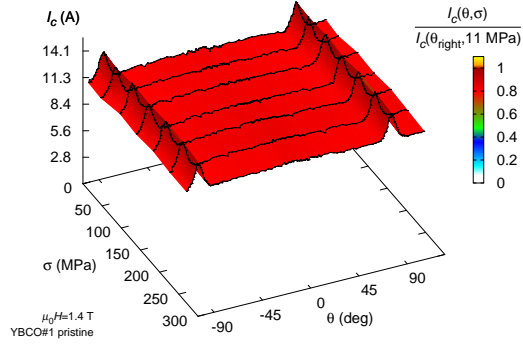}
		\caption{\label{AnisotropyIcYBCO}Critical current anisotropy of the pristine YBCO/RABiTS sample at \muH=1.4\,T.}
	\end{center}
\end{figure}

 Figure \ref{AnisotropyIcYBCO}\footnote{Note that the color code in all following 3D-figures refers to a normalization of $I_\mathrm{c}$ by the (right) maximum near \Hpab (at 90\deg\,and around 85\deg\,in the YBCO/RABiTS and GdBCO/IBAD tape, respectively. 0\deg\,refers to the field parallel to the c-axis.)} shows the dependence of $I_\mathrm{c}$ in the pristine YBCO/RABiTS sample on the applied stress at \muH=1.4\,T.

The two maxima of $I_\mathrm{c}$ occur for fields parallel to the a,b-planes and the $I_\mathrm{c}$-anisotropy\,(ratio of $I_c($\HpabTemp$)/I_c($\HpcTemp$)$) is quite low. As expected from the self-field measurements, the applied stress hardly influences the critical current up to the maximum applied stress of 303\,MPa. For lower fields\,(not shown), the anisotropy is even smaller and no significant influence of applied stress was found either.

\begin{figure}[htpb]
	\centering
	\begin{center}
		\includegraphics[width=7cm]{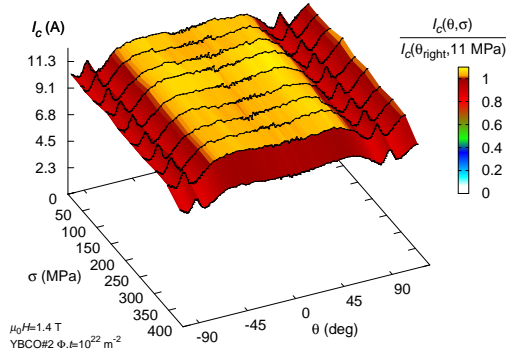}
		\caption{\label{AnisotropyIcYBCOIrradiated}Critical current anisotropy of the irradiated YBCO/RABiTS sample at \muH=1.4\,T.}
	\end{center}
\end{figure}

\begin{figure}[htpb]
	\centering
	\begin{center}
		\includegraphics[width=7cm]{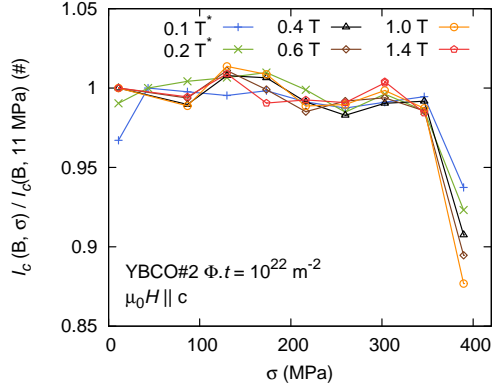}
		\caption{\label{cPeakYBCO2}Stress dependence of $I_\mathrm{c}$ of the irradiated YBCO/RABiTS sample for different fields \HpcTemp. Note that the 0.1\,T and 0.2\,T values are normalized by the values at 44\,MPa as the temperature was too high at 0.11\,MPa.}
	\end{center}
\end{figure}

As already discussed, $I_\mathrm{c}$ of the YBCO/RABiTS tape is enhanced by the irradiation between the a,b-peaks\,(figure \ref{AnisotropyIcYBCOIrradiated}). 
The a,b-peaks on the other hand are reduced.
No stress dependence of the in-field $I_\mathrm{c}$ in the irradiated sample was found within experimental accuracy up to the second highest applied stress of 346\,MPa, but a reversible degradation to 90\% of the initial $I_\mathrm{c}$ occurs at 389\,MPa. Furthermore, no field dependence of the stress-sensitivity was observed up to 346\,MPa for all orientations. Figure \ref{cPeakYBCO2} shows the $I_\mathrm{c}$-dependence on the tensile stress at various fields\,(\HpcTemp). At the highest applied stress of 389\,MPa, the degradation is higher at higher fields. This effect is less pronounced at \HpabTemp (not shown). 

Contrary to the YBCO/RABiTS tape, $I_\mathrm{c}$ of the pristine sample is asymmetric with respect to the c-axis in the GdBCO/IBAD conductor. At 0.1\,T, the peaks are shifted by -11\deg\,($\Theta_{left}$) and -6.5\deg \,($\Theta_{right}$) respectively with respect to the tape surface. With increasing field the shift is reduced to about -4\deg at 1.4\,T for both peaks. Furthermore, $I_\mathrm{c}$ at $\Theta_{right}$ is always higher than at $\Theta_{left}$. A more detailed description of the angular dependence in similar tapes can be found elsewhere \cite{MaiorovAPL86,ChenAPL94,ChudySUST24}.
The applied tensile stress neither influences the angular dependence of $I_\mathrm{c}$ nor the positions of the local maxima.

Note that a reduction of the peak shift from 12\deg ~towards 0\deg ~as well as a change of the angular $I_\mathrm{c}$-dependence were observed under \textit{compressive} strain at 0.5\,T in Ref. \cite{HigginsIEEE21} in an YBCO/IBAD sample.

\begin{figure}[htpb]
	\centering
	\begin{center}
		\includegraphics[width=7cm]{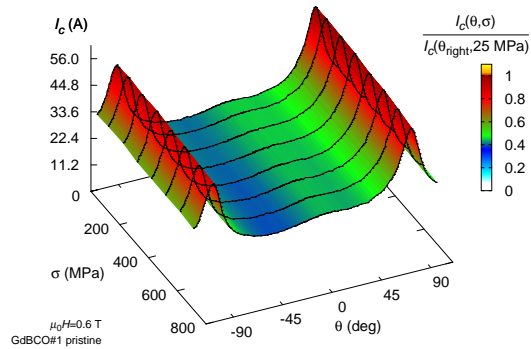}
		\caption{\label{AnisotropyIcGdBCO}Critical current anisotropy of the pristine GdBCO/IBAD sample at \muH=0.6\,T.}
	\end{center}
\end{figure}

\begin{figure}[htpb]
	\centering
	\begin{center}
		\includegraphics[width=7cm]{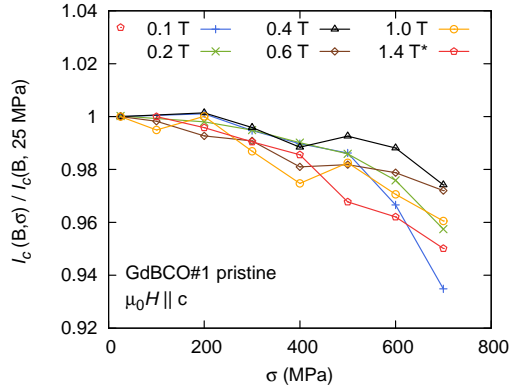}
		\caption{\label{cPeakGdBCO1unirr}Stress dependence of $I_\mathrm{c}$ for \Hpc of the pristine GdBCO/IBAD sample at different fields. At 25\,MPa/1.4\,T the temperature was too high. Therefore, $I_\mathrm{c}$ of the 1.4\,T measurement was normalized by $I_\mathrm{c}$ at 100\,MPa}
	\end{center}
\end{figure}

Figure \ref{AnisotropyIcGdBCO} shows the angular and stress dependence of $I_\mathrm{c}$ in the pristine sample at \muH=0.6\,T. $\Theta_{left}$ corresponds to the local peak at around -95\deg, $\Theta_{right}$ to the local peak at around 85\deg.
A local minimum occurs at around -30\deg. As in the self-field measurements, the critical current does not significantly change with the applied stress in the reversible region.
The field and stress dependence on the normalized $I_\mathrm{c}$ for \Hpc is shown in figure \ref{cPeakGdBCO1unirr} for the pristine sample. 
The reduction of $I_\mathrm{c}$ is weaker at 0.4\,T, especially at higher applied stress. A weaker stress sensitivity of $I_\mathrm{c}$ in YBCO/IBAD samples at around \muH=0.2\,T-0.25\,T was also found in \cite{SuganoSUST21}\cite{VanDerLaanSUST23}. This field dependence of the stress sensitivity is accompanied by a shift of the maximum $I_\mathrm{c}$ from zero to finite strain in YBCO/IBAD samples \cite{VanDerLaanSUST23}\cite{SuganoSUST21}, which is not observed in the GdBCO/IBAD samples. 

For $H \| \Theta_{right}$ $I_\mathrm{c}$ decreases monotonously with increasing field\,(not shown) like in the YBCO/MOD sample, but the reduction starts already at lower applied stress.

After irradiation, the angular dependence of $I_\mathrm{c}$ becomes more symmetric with respect to the c-axis and both maxima are located approximately -7\deg ~off the tape surface at 0.1\,T. The maxima shift with field and converge at -4\deg ~at 1.4\,T. The reduction of $I_\mathrm{c}$ by stress is significantly higher over the entire angular range compared to the pristine sample. 

\begin{figure}[htpb]
	\centering
	\begin{center}
		\includegraphics[width=7cm]{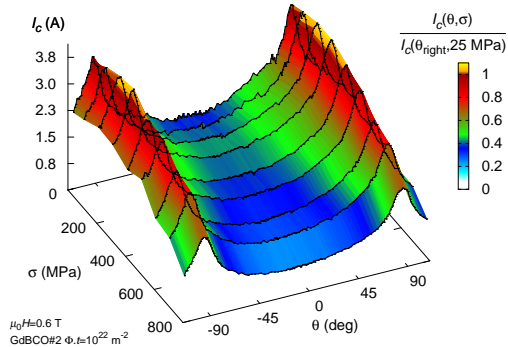}
		\caption{\label{AnisotropyIcGdBCOIrradiated}Critical current anisotropy of the irradiated GdBCO/IBAD sample at \muH=0.6\,T.}
	\end{center}
\end{figure}

Figure \ref{AnisotropyIcGdBCOIrradiated} shows the angular and stress dependence of $I_\mathrm{c}$ at 0.6\,T.
A clear maximum at around 200\,MPa occurs over the entire angular range. The minimum of $I_\mathrm{c}$ is found at around \Hpc independently of stress. 

\begin{figure}[htpb]
	\centering
	\begin{center}
		\includegraphics[width=7cm]{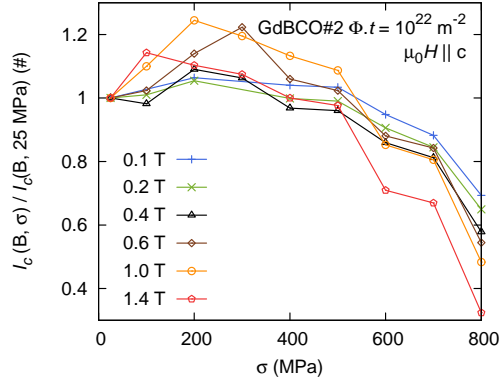}
		\caption{\label{cPeakGdBCO1}Stress dependence of $I_\mathrm{c}$ for \Hpc of the irradiated GdBCO sample for different fields.}
	\end{center}
\end{figure}

The field and stress dependent change of $I_\mathrm{c}$ at \Hpc is illustrated in figure \ref{cPeakGdBCO1}.
The critical current reaches a maximum at an applied stress of 200\,-300\,MPa. At intermediate fields (0.6\,T and 1.0\,T), significantly higher values of $I_\mathrm{c}$ are reached at 200\,MPa. On the other hand, at higher applied stress, the reduction of the critical currents increases monotonously with field. 

At $H \| \Theta_{right}$ the maximum of $I_\mathrm{c}$ at around 200\,MPa was not as pronounced as for \Hpc, but still visible. At high applied stress, $I_\mathrm{c}$ is once again more strongly reduced at higher applied field. 

\section{Conclusion}
Measurements of the critical currents under tensile stress in pristine and neutron irradiated YBCO/MOD and GdBCO/IBAD coated conductors were performed in magnetic fields of up to 1.4\,T. The field orientation varied within a range of about 220\deg ~including both a,b-peaks\,(\HpabTemp).

The critical currents in GdBCO/IBAD were significantly reduced after irradiation due to the strong reduction of $T_\mathrm{c}$ by 6.2\,K. Thermal and epi-thermal neutrons captured by the $^{155}$Gd and $^{157}$Gd-nuclei seem to be responsible for this strong reduction, as the recoil energy during gamma emission introduces additional point defects. These point defects are responsible for enhanced disorder. Thermal neutron absorption does not play a role in YBCO because the neutron absorption cross sections are too small for all constituents of YBCO.

The irreversible stress/strain limit of the GdBCO/IBAD and the YBCO/MOD sample do not change significantly after irradiation. The stress dependence of $I_\mathrm{c}$ in YBCO/MOD did not change after irradiation, whereas the sensitivity of $I_\mathrm{c}$ to stress was significantly enhanced in GdBCO/IBAD. This change can be ascribed to the lower $T_\mathrm{c}$ and the resulting higher ratio of $T/T_c$ at liquid nitrogen temperature. At the highest applied field\,(1.4\,T), $I_\mathrm{c}$ at 800\,MPa was reduced to less than 40\% of the relaxed $I_\mathrm{c}$-value. 

Further experiments, where the thermal neutrons are shielded during the irradiation, have to complement the existing data to judge the possible application of GdBCO conductors in future fusion devices.
\ack
The authors thank American Superconductor and SuperPower for providing the samples used in this study.
Furthermore, we thank Ventsislav Mishev, Florian Hengstberger, Herbert Hartmann and Rudolf Gergen for important technical support.

This work was supported in part by the Friedrich Schiedel Foundation for Energy Technology, Vienna, and by the European Commission under the Contract of Association between EURATOM and the Austrian Academy of Sciences. It was carried out within the framework of the European Fusion Development Agreement. The views and opinions expressed herein do not necessarily reflect those of the European Commission.

\section*{References}
\bibliographystyle{unsrt}
\bibliography{paper}
\end{document}